\documentclass[11pt,a4paper]{article}

\usepackage[english]{babel}
\usepackage{authblk}
\usepackage{graphicx}
\usepackage{hyperref} 
\usepackage{geometry}
\usepackage{amsmath}
\usepackage{amssymb}
\usepackage{amsfonts}
\usepackage{amsthm}
\usepackage{multicol}
\usepackage{slashed}
\usepackage{float}
\usepackage{caption}
\usepackage{cite}
\usepackage{xcolor}
\usepackage{bbm}
\numberwithin{equation}{section}

\newtheorem*{mydef}{Definition}

\newcommand{\ri}{{\rm i}}
\newcommand{\rd}{{\rm d}}

\newcommand{\R}{\mathbb{R}}
\newcommand{\E}{\mathbb{E}}
\newcommand{\p}{\partial}
\newcommand{\indep}{\perp \!\!\! \perp}

\title{\bf A Stochastic Origin of Spacetime Non-Commutativity}

\author[1,2]{Michele Arzano\thanks{Email: michele.arzano@na.infn.it}}
\author[2,3]{Folkert~Kuipers\thanks{Email: f.kuipers@physik.uni-muenchen.de}}

\affil[1]{\em Dipartimento di Fisica ``E. Pancini", Universit\`a di Napoli Federico II, I-80125 Napoli, Italy\\}

\affil[2]{\em INFN, Sezione di Napoli\authorcr \em Complesso Universitario di Monte S. Angelo\authorcr \em Via Cintia Edificio 6, 80126 Napoli, Italy}

\affil[3]{\em Arnold Sommerfeld Center for Theoretical Physics\\
Ludwig-Maximilians-Universit\"at M\"unchen, 80333 M\"unchen, Germany}

\begin{document}

\maketitle

\begin{abstract}
	We propose a stochastic interpretation of spacetime non-commutativity starting from the path integral formulation of quantum mechanical commutation relations. We discuss how the (non-)commutativity of spacetime is inherently related to the continuity or discontinuity of paths in the path integral formulation. Utilizing Wiener processes, we demonstrate that continuous paths lead to commutative spacetime, whereas discontinuous paths correspond to non-commutative spacetime structures. As an example we introduce discontinuous paths from which the $\kappa$-Minkowski spacetime commutators can be obtained. Moreover we focus on modifications of the Leibniz rule for differentials acting on discontinuous trajectories. We show how these can be related to the deformed action of translation generators focusing, as a working example, on the $\kappa$-Poincar\'e algebra. Our findings suggest that spacetime non-commutativity can be understood as a result of fundamental discreteness in temporal and/or spatial evolution. 
\end{abstract}

\section{Introduction}

Space-time non-commutativity gained significant popularity in the late 1990s, primarily due to its relevance in certain limits of string theory involving a constant background B-field \cite{Sheikh-Jabbari:1997qke,Seiberg:1999vs, Connes:1997cr}. As already noted by the authors of \cite{Seiberg:1999vs}, the idea was not new; indeed, a model of ``quantized space-time" in which ordinary coordinates are replaced by non-commuting hermitian operators had already appeared fifty years earlier in the seminal work of Snyder \cite{Snyder:1946qz}. The main motivation of this work was that non-commutativity of coordinates, through the introduction of a fundamental length, could establish a natural cut-off scale in quantum field theory, potentially resolving the issues with infinities that were prevalent at the time.\footnote{Interestingly, as reported in \cite{Jackiw:2001dj}, the suggestion that non-commuting space-time coordinates could be used to tame divergences in quantum field theory might have been originally suggested by Heisenberg.}

In an almost parallel development Doplicher, Fredenhagen, and Roberts (DFR) explored the possible role of a non-commutative, ``quantized space-time", for physics at the Planck scale \cite{Doplicher:1994zv, Doplicher:1994tu}. These authors suggested that non-trivial commutation relations between coordinates can be motivated by considerations involving gedanken experiments combining Heisenberg's uncertainty principle and classical gravity which result in limitations to localization in space-time.

In both the DFR model and the stringy scenario one deals with a commutator of space-time coordinates given by a matrix of constants. This raises the issue of breaking of Lorentz symmetry arising from the very structure of the commutator, since, at first glance, the commutator itself is not invariant under Lorentz transformations. In contrast in the model proposed by Snyder the space-time commutators transform covariantly under Lorentz transformations at the price of embedding the space-time coordinate operators into a larger algebra. In the work \cite{Majid:1994cy}, it was shown that one can bypass both shortcomings and introduce space-time non-commutativity with a structure that is covariant under a {\it deformation} of the Poincar\'e group, based on the theory of Hopf algebras and quantum groups \cite{Lukierski:1991pn,Lukierski:1992dt}. This connection further boosted interest in the study of non-commutative space-times as models of a ``flat space-time limit" of quantum gravity \cite{Amelino-Camelia:2003ezw,Freidel:2003sp}, in which the Planck energy/length plays the role of a second observer-independent scale, thanks to the non-trivial structures of the deformed Poincar\'e group \cite{Amelino-Camelia:2000cpa,Amelino-Camelia:2000stu}. Models of space-time non-commutativity are now among the most studied in the context of quantum gravity phenomenology, a fast-growing field of research looking for experimental signatures of the putative quantum features of spacetime \cite{Amelino-Camelia:2008aez, Addazi:2021xuf}.

There is a vast body of literature exploring various physical and geometrical consequences of the non-commutativity of space-time. In specific models where the commutator of space-time coordinates forms a Lie algebra, the associated four-momentum space is found to be a Lie group \cite{Arzano:2010jw,Arzano:2014ppa}. This aspect is closely related to the Hopf algebra deformation of the Poincar\'e group \cite{Arzano:2022ewc}. Another interesting aspect is the potential connection between spacetime non-commutativity and discrete features of spacetime. For a specific model of Lie-algebra-type non-commutativity, early indications of an underlying discrete structure were presented in \cite{Amelino-Camelia:1999jfz}, where it was suggested that to consistently define a differential calculus, derivatives should be replaced by finite differences. In \cite{Amelino-Camelia:2017utp}, it was proposed that deformations of the Poincar\'e algebra, compatible with the same type of Lie-algebra non-commutativity, might be linked to discrete features induced by polymer quantization, a procedure that encapsulates key aspects of loop quantum gravity. More recently \cite{Lizzi:2021dud}, it was shown that the $\rho$-Minkowski non-commutative model requires the spectrum of time to be discrete. 


This work aims to deepen our understanding of the potential discrete features underlying space-time non-commutativity by providing a complementary perspective to the result presented in \cite{Lizzi:2021dud}. 
Specifically, we argue that space-time non-commutativity can be viewed as a consequence of a fundamental discrete and stochastic structure of space-time. Our main inspiration is the stochastic interpretation of quantum mechanics \cite{Feynman:1948ur,Nelson:1966sp,Nelson:1985,Guerra:1981ie,Pavon:1995April,Pavon:2000,Kuipers:2023pzm,Kuipers:2023ibv} in which the canonical commutator between position and momentum operators can be obtained imposing non-differentiability of the paths in Feynman's path integral. We argue that if one replaces such non-differentiable paths
with discontinous stochastic processes then coordinates can acquire a non trivial commutator and we show a specific example of such paths which reproduces the much studied $\kappa$-Minkowski non-commutative space-time. This is done in section \ref{sec:PathIntegral}, where we also recall the basics of the stochastic approach to quantum mechanics and how the canonical commutator can be obtained from the path integral.
First, in section \ref{sec:PathIntegral}, we will provide an argument for this statement in a path integral approach. Then, in section \ref{sec:Leibniz}, we provide further proof for this statement by considering modifications of Leibniz' rule. Finally, in section \ref{Conclusion}, we provide concluding remarks and an outlook to future developments. The appendix \ref{ap:Wiener} reviews some elementary aspects of the Wiener process that will be used in section \ref{sec:PathIntegral}.

\section{Non-commutative spacetime from fundamental discreteness: the path integral approach}\label{sec:PathIntegral}
The aim of this section is to show how one can use a path integral approach to argue that spacetime non-commutativity is a consequence of a fundamental discreteness of temporal and/or spatial evolution. As a first step, in section \ref{sec:Ordering}, we recall the relation between operator ordering in canonical quantization and path ordering in the path integral formalism \cite{Feynman:1948ur}. We will then proceed, in section \ref{sec:PIvsWP}, by reviewing the relation between path integrals and the Wiener process. This allows to derive phase space non-commutativity in the canonical approach from the non-differentiability of the paths in the path integral, which is done in section \ref{sec:PhaseSpaceNC}. Finally, in section \ref{sec:NonContPath}, we will generalize this derivation to argue that spacetime non-commutativity arises when the path integral integrates over discontinuous paths.

\subsection{Operator ordering vs. time ordering}\label{sec:Ordering}
We consider the classical trajectory of a particle in Euclidean space $X:\mathcal{T}\rightarrow\R^{n}$ with $\mathcal{T}=[0,T]\subset\R$. The trajectory minimizes the action given by
\begin{equation}
	S(X) = \int_\mathcal{T} \frac{m}{2} \, \delta_{ij} \frac{dX^i}{dt} \frac{dX^j}{dt}  \, dt \, . 
\end{equation}
Let us now consider two points along the trajectory $x_0 = X(0)$ and $x_T = X(T)$. In the quantum version of the model, the transition between these points is described by a probability amplitude determined by a wave function $\Psi(x,t)$, such that $\Psi(x,0)=\Psi_0(x)$ and $\Psi(x,T)=\Psi_T(x)$. Then, any state $\Psi(x,t)$ with $t\in\mathcal{T}$ can be obtained from the initial state $\Psi(x,0)$ by applying the time evolution operator, i.e.
\begin{equation}\label{eq:PTimeEvo}
	|\Psi(x,t) \rangle = e^{ - \ri \, \hat{H} \, t} \, |\Psi_0(x)\rangle \, .
\end{equation}
where $\hat{H}$ is the Hamiltonian operator.
This time evolution is governed by the Schr\"odinger equation 
\begin{equation}
	\ri \, \hbar \, \frac{\p}{\p t} \Psi(x,t) = - \frac{\hbar^2}{2 \, m} \, \delta^{ij} \p_j \p_i \Psi(x,t)\, .
\end{equation}

Let us discuss how one can describe this time evolution in the path integral approach. In this case, the probability amplitude for evolving from the event $x_0$ to $x_T$ is given by the integral over the paths
\begin{equation}
	\langle x_T, T \, | \, x_0, 0 \rangle 
	= \int_{X(0)=x_0}^{X(T)=x_T} DX(t) \, e^{\frac{\ri \, S[X(t)]}{\hbar}}\, .
\end{equation}
By inserting a complete set of states at an arbitrary time $t\in(0,T)$, the left hand side can be rewritten as
\begin{equation}
	\langle x_T, T \, | \, x_0, 0 \rangle 
	=
	\int dx  \, \langle x_T, T \, | x,t \rangle \langle x,t | \, x_0, 0 \rangle \, ,
\end{equation}
such that
\begin{align}
	\langle x_T, T \, | \, x_0, 0 \rangle
	=
	\int dx 
	\int_{X(t)=x}^{X(T)=x_T} DX(s) \, e^{\frac{\ri \, S[X(s)]}{\hbar}}
	\int_{X(0)=x_0}^{X(t)=x} DX(r) \, e^{\frac{\ri \, S[X(r)]}{\hbar}} \, .
\end{align}
This split property allows to calculate the matrix element for the operator $\hat{x}$ in the path integral formulation as\footnote{Cf. e.g. Appendix A in Ref.~\cite{Polchinski:1998rq}.}
\begin{align}\label{eq:MatrixElem}
	\langle x_T, T \,|\, \hat{x}(t) \,|\, x_0, 0 \rangle 
	&=
	\int dx \, \langle x_T, T \, | \, \hat{x}(t)\, |\, x,t \rangle \langle x,t | \, x_0, 0 \rangle 
	\nonumber\\
	&=
	\int dx \, \langle x_T, T \, | \, x,t \rangle\, x \,\langle x,t | \, x_0, 0 \rangle 
	\nonumber\\
	&=
	\int_{X(0)=x_0}^{X(T)=x_T} DX(t) \, e^{\frac{\ri \, S[X(t)]}{\hbar}} \, X(t) \, ,
\end{align}
where, in the first equality, we inserted the integral over a complete set of position eigenstates $ |\, x,t \rangle$ at the time $t$.
One can generalize this by considering a path integral for the product of two operators $\hat{A}$ and $\hat{B}$. This yields
\begin{equation}
	\langle x_T, T \,|\, T[\hat{A}(t) \, \hat{B}(t') ] \,|\, x_0, 0 \rangle 
	=
	\int_{X(0)=x_0}^{X(T)=x_T} DX(t) \, e^{\frac{\ri \, S[X(t)]}{\hbar}} \, A(t) B(t') \, ,
\end{equation}
where $T[\hat{A}(t) \, \hat{B}(t') ]$ denotes the time ordered product of two operators, defined as
\begin{equation}
	T[\hat{A}(t) \, \hat{B}(t') ]
	:=
	\theta(t-t') \, \hat{A}(t) \hat{B}(t') + \theta(t'-t) \hat{B}(t') \hat{A}(t) \, .
\end{equation}
In the limit $t'\rightarrow t$ this product becomes ambiguous for non-commuting operators, as the limit depends on whether it is taken from above or below:
\begin{align}
	\lim_{t'\rightarrow t^+} &= \hat{B}(t) \hat{A}(t)\, ,\\
	\lim_{t'\rightarrow t^-} &= \hat{A}(t) \hat{B}(t) \, .
\end{align}
Therefore, the commutator of $\hat{A}$ and $\hat{B}$ can be written in terms of the time ordered product as
\begin{align}
	[\hat{A},\hat{B}](t) &= \lim_{t'\rightarrow t^-} T[\hat{A}(t) \, \hat{B}(t') ] - \lim_{t'\rightarrow t^+} T[\hat{A}(t) \, \hat{B}(t') ] \nonumber\\
	&= \lim_{\epsilon\rightarrow 0^+} T[\hat{A}(t) \, \hat{B}(t-\epsilon)] - T[\hat{A}(t) \, \hat{B}(t+\epsilon)] \, .
\end{align}
Thus, the commutator of two operators can be written in terms of a path integral as 
\begin{align}\label{eq:OrderingCorrespondence}
	\langle x_T, T \,|\, [\hat{A},\hat{B}](t) \,|\, x_0, 0 \rangle 
	&=
	\lim_{\epsilon\rightarrow 0^+}\int_{X(0)=x_0}^{X(T)=x_T} DX(t) \, e^{\frac{\ri \, S[X(t)]}{\hbar}} 
	\nonumber \\
	&\qquad \times 
	\Big\{ A(t) \, B(t-\epsilon)
	-  B(t+\epsilon) \, A(t) \Big\}.
\end{align}
Hence, the operator ordering in the canonical quantization approach can be understood in terms of a time ordering in the path integral formulation \cite{Feynman:1948ur}.

\subsection{Path integrals and Wiener processes}\label{sec:PIvsWP}
Before discussing explicit examples of non-commutative variables in a path integral formulation, we recall some properties of path integrals. Our starting point is the path integral expression \eqref{eq:MatrixElem} for an arbitrary operator $\hat{A}$:
\begin{equation*}
	\langle x_T, T \,|\, \hat{A}(t) \,|\, x_0, 0 \rangle 
	=
	\int_{X(0)=x_0}^{X(T)=x_T} DX(t) \, e^{ \frac{\ri}{\hbar} \, S[X(t)] } \, A(t) \, .
\end{equation*}
Heuristically, this path integral sums over an appropriate set of paths and weighs every path with their probability of occurring determined by the measure on the space of all paths 
\begin{equation}\label{eq:LorMeasure}
	\rd\mu(X)=e^{\frac{\ri \, S[X]}{\hbar}} DX \, .
\end{equation}
Although this heuristic definition is intuitively clear, a mathematically rigorous construction of this integral for quantum systems has remained a vexing problem, ever since the introduction of the path integral, cf. e.g. \cite{Albeverio}. The problems in defining a path integral are much less severe in the Euclidean formulation. In this case, a Wick rotation \cite{Wick:1954eu} is applied, such that the path integral becomes
\begin{equation}\label{eq:EuclPathInt}
	\langle x_T, T \,|\, \hat{A}(t) \,|\, x_0, 0 \rangle 
	=
	\int_{X(0)=x_0}^{X(T)=x_T} DX(t) \, e^{- \frac{S[X(t)]}{\hbar} \, } \, A(t)
\end{equation}
with measure
\begin{equation}\label{eq:EucLMeasure}
	\rd\mu(X)=e^{\frac{- \, S[X]}{\hbar}} DX \, .
\end{equation}
This integral can be defined rigorously as the measure can be related to the Wiener measure \footnote{This has served as a starting point for the Euclidean approach to quantum field theory \cite{Schwinger:1958mma,NelsonPath,Osterwalder:1973dx,Osterwalder:1974tc,Glimm:1987ng}.} \cite{Kac:1949}, which is associated to the Wiener integral \cite{Wiener}.
\par 

It follows that, in the Euclidean formulation, the classical trajectory $X:\mathcal{T}\rightarrow\R^{n}$ is promoted to a stochastic process $X:\mathcal{T}\times\Omega\rightarrow\R^{n}$, where $\Omega$ is called a sample space. Its elements $\omega\in\Omega$ label the sample paths $X(\cdot,\omega): \mathcal{T}\rightarrow\R^{n}$, which are precisely the paths that the path integral sums over. Moreover, these paths are weighted by a probability measure $\mathbb{P}$, which assigns a probability to every $\omega\in\Omega$. This probability measure is defined in such a way that the induced measure on the path space, i.e. $\mu_X=\mathbb{P}\circ X^{-1}$, satisfies eq.~\eqref{eq:EucLMeasure}. Since the Wiener measure is generated by a Wiener process \cite{Doob:1942}, one finds that compatibility with \eqref{eq:EucLMeasure} imposes the process $X$ to be a Wiener process with drift and diffusion constant $\hbar/m$ \cite{Kac:1949,Albeverio}.
Thus, $X$ can be decomposed as
\begin{equation}\label{eq:DoobMeyer}
	X = C + \sqrt{\frac{\hbar}{m}} \, W \, ,
\end{equation}
where $C$ represents the drift, that is determined by the stationary action principle, and $W$ is a standard Wiener process, which is reviewed in appendix \ref{ap:Wiener}. This process inherits various properties of the Wiener process. Examples of such properties are \cite{Morters:2010}:
\begin{itemize}
	\item The sample paths of a Wiener process are continuous everywhere, but nowhere differentiable. More precisely, the sample paths of a Wiener process are $\alpha$-H\"older continuous\footnote{Cf. section \ref{sec:ContLeibniz} for a definition of H\"older continuity.} for $\alpha\in(0,\frac{1}{2})$, but fail to be $\alpha$-H\"older continuous for any $\alpha\geq\frac{1}{2}$.
	\item The Hausdorff dimension of a Wiener process is ${\rm dim}_{H} \{ X(t,\omega) \, | \, t\in\mathcal{T}, \, \omega\in\Omega \}=2$. In contrast, for the classical trajectory one finds ${\rm dim}_{H}\{X(t)\,|\, t\in\mathcal{T}\}=1$.
\end{itemize}
It follows that the paths in the Wick-rotated path integral obey the same properties. In the remainder of this paper, we will focus on the first property of non-differentiability of these paths.\footnote{Note that this fact was known to Feynman, as illustrated by the following quote: ``The paths involved are, therefore, continuous but possess no derivative.
	They are of a type familiar from study of Brownian motion.'' \cite{Feynman:1948ur}.}
\par 

We conclude this subsection by emphasizing that our arguments rely on the use of a Wick rotation, as the above properties have only been proven for the Wick rotated path integral. It has been suggested \cite{Gelfand} that the original (non Wick-rotated\footnote{In relativistic theories, this would be called the Lorentzian path integral. However, here we work with a non-relativistic theory, such that the signature is Euclidean, both before and after applying the Wick rotation. In the non-relativistic case, the Wick rotation only affects the complex factor in the exponent of the path integral measure, whereas, in the relativistic case, it affects both the spacetime signature and the complex factor.}) path integral could be constructed in a similar way by considering a stochastic process $X:\mathcal{T}\times\Omega\rightarrow\R^{n}$ with complex diffusion constant $\ri \, \hbar/m$, such that the induced measure, $\mu=\mathbb{P}\circ X^{-1}$, satisfies eq.~\eqref{eq:LorMeasure}. However, it has been shown that there does not exist such a probability measure $\mathbb{P}$ for real paths $X$ \cite{Cameron,Daletskii}, as the existence requires the diffusion constant to be positive definite. A solution to this problem for the single particle system has been proposed in Refs.~\cite{Kuipers:2023ibv,Kuipers:2023pzm}. The idea is to consider complex paths $Z:\mathcal{T}\times \Omega\rightarrow\mathbb{C}^{n}$ with diffusion constant $\ri \, \hbar/m$. For this diffusion constant, $X={\rm Re}(Z)$ and $Y={\rm Im}(Z)$ are maximally correlated Wiener processes with drift. In contrast to the real case, the probability measure exists for this complex process. Moreover, the real projection $X=\mathcal{T}\times \Omega\rightarrow\R^{n}$ of these complex paths can be identified with the paths appearing in the quantum path integral \cite{Pavon:2000}. Using this construction, all arguments presented in this paper can be extended to the case of the non-Wick rotated path integral.

\subsection{Canonical commutation relation from non-differentiable paths}\label{sec:PhaseSpaceNC}
%
Using the properties of the Wiener process, we can derive the canonical commutation relation of quantum mechanics from the non-differentiability of the paths in the path integral. 

In the derivation that follows, we will make use of various notions of velocity that can be associated to a Wiener process. The first two are the forward and backward velocity given by the limits
\begin{align}
	V_+^i(t) &\sim \lim_{\epsilon\rightarrow 0} \frac{X^i(t+\epsilon) - X^i(t)}{\epsilon} \, ,\nonumber\\
	V_-^i(t) &\sim \lim_{\epsilon\rightarrow 0} \frac{X^i(t) - X^i(t-\epsilon)}{\epsilon} \, ,
\end{align}
where we use the $\sim$ symbol to indicate that these are heuristic definitions. In order to properly construct these velocities, one requires conditional expectations and the It\^o integral, cf. appendix \ref{ap:WPV}. In these expressions, the non-differentiability of the Wiener process is reflected in the fact that, in general, $V_+\neq V_-$. In addition, we require a notion of second order velocity defined by
\begin{equation}
	V_2^{ij}(\tau) = \lim_{\epsilon\rightarrow 0} \frac{[X^i(t+\epsilon) - X^i(t)] [X^j(t+\epsilon) - X^j(t)]}{\epsilon} \, .
\end{equation}
This is the velocity associated to the quadratic variation of the Wiener process, cf. appendix \ref{sec:QVar}. It measures the velocity of the variance\footnote{More precisely, $V_\pm$ describe a velocity associated to the expectation value (first moment), whereas $V_2$ describes the velocity of the second moment.} of the process and has physical dimension $[V_2]=L^2/T$. It can easily be checked, by means of a Taylor expansion, that for differentiable paths this velocity vanishes. However, it may be non-vanishing for non-differentiable paths. A typical example of such paths are the sample paths of the Wiener process, as discussed in section \ref{sec:PIvsWP}. For the process \eqref{eq:DoobMeyer}, this velocity is given by,\footnote{In fact, this relation can be taken as one of the defining properties of the Wiener process \cite{Levy:1948}.} cf. eq.~\eqref{eq:DoobMeyer} and Appendix \ref{sec:QVar},
\begin{equation}\label{VelQVarWien}
	V_2^{ij}(t) = \frac{\hbar}{m} \, \delta^{ij} \, .
\end{equation}
\par 

Using the properties of the Wiener process, we can now derive the canonical commutation relation in Feynman's path integral approach \cite{Feynman:1948ur}. In order to do so, we consider the commutator $[\hat{x}^i,\hat{p}_j]$ and use eq.~\eqref{eq:OrderingCorrespondence} to write it in terms of the path integral of a time-ordered difference:
\begin{align}\label{eq:CanComrel1}
	\langle x_T, T \,|\, [\hat{x}^i,\hat{p}_j](t) \,|\, x_0, 0 \rangle 
	&=
	\lim_{\epsilon\rightarrow 0^+}\int_{X(0)=x_0}^{X(T)=x_T} DX(t) \, e^{- \, \frac{S[X(t)]}{\hbar}} 
	\nonumber \\
	&\qquad \times 
	\Big\{ X^i(t) \, P_j(t-\epsilon)
	-  P_j(t+\epsilon) \, X^i(t) \Big\}.
\end{align}
In the Euclidean theory $X$ and $P$ are the position and momentum of the Wiener process, as discussed in section \ref{sec:PIvsWP}. Thus, in order to obtain the commutation relation in the Euclidean theory, we must evaluate this expression for a Wiener process. Using its canonical expression, the momentum can be written as a velocity
\begin{equation}
	P_i = \frac{\p L(X,V)}{\p v^i} = m \, \delta_{ij} \, V^j \, ,
\end{equation}
and eq.~\eqref{eq:CanComrel1} can be rewritten as 
\begin{align}
	\langle x_T, T \,|\, [\hat{x}^i,\hat{p}_j](t) \,|\, x_0, 0 \rangle 
	&=
	\int_{X(0)=x_0}^{X(T)=x_T} DX(t) \, e^{ - \frac{S[X(t)] }{\hbar}} 
	m \, \delta_{jk} \Big\{ X^i(t) \, V_-^k(t)
	-  V^k_+(t) \, X^i(t) \Big\}.
\end{align}
As the velocity processes are only well-defined with respect to the integral (cf. appendix \ref{ap:WPV}) we will integrate this expression over an arbitrary interval $I\subseteq \mathcal{T}$. This yields 
\begin{align}
	\left\langle x_T, T \,\left|\, \int_I[\hat{x}^i,\hat{p}_j](t) \, dt \,\right|\, x_0, 0 \right\rangle 
	&=
	m \, \delta_{jk} \int_{X(0)=x_0}^{X(T)=x_T} DX(t) \, e^{- \frac{S[X(t)]}{\hbar} \, } 
	\nonumber\\
	&\qquad \times 
	\int_I \Big\{ X^i(t) \, V_-^k(t)
	-  V^k_+(t) \, X^i(t) \Big\} dt
	\nonumber\\
	&=
	m \, \delta_{jk} \int_{X(0)=x_0}^{X(T)=x_T} DX(\tau) \, e^{- \frac{S[X(t)]}{\hbar}} 
	\nonumber\\
	&\qquad \times 
	\int_I \left[  X^i(t) \, \rd_-X^k(t) - X^i(t) \, \rd_+ X^k(t) \right] , 
\end{align}
where we used eq.~\eqref{eq:IntDefVel} in the second line. Moreover, the It\^o integrals $\int \rd_\pm X$ are defined in eqs.~\eqref{eq:ItoFDef} and \eqref{eq:ItoBDef}. Using eqs. \eqref{eq:Vel2nd} and \eqref{eq:QvarvsIto}, this expression can be rewritten in terms of an integral over quadratic variation as
\begin{align}
	\left\langle x_T, T \,\left|\, \int_I [\hat{x}^i,\hat{p}_j](t) \, dt \,\right|\, x_0, 0 \right\rangle 
	&=
	m \, \delta_{jk} \int_{X(0)=x_0}^{X(T)=x_T} DX(t) \, e^{- \, \frac{S[X(t)]}{\hbar}} \int_I \, \rd[X^i,X^k](t) 
	\nonumber\\
	&=
	m \, \delta_{jk} \int_{X(0)=x_0}^{X(T)=x_T} DX(t) \, e^{- \, \frac{S[X(t)]}{\hbar}} \int_I V_2^{ik}(t) \, dt \, .
\end{align}
Then, after removing the integral on both sides and plugging in the expression for the quadratic variation of the Wiener process \eqref{VelQVarWien}, one finds
\begin{align}
	\left\langle x_T, T \left| [\hat{x}^i,\hat{p}_j](t) \right| x_0, 0 \right\rangle 
	&=
	\int_{X(0)=x_0}^{X(T)=x_T} DX(t) \, e^{- \frac{S[X(t)]}{\hbar}}  m \, \delta_{jk} \, V_2^{ik} \nonumber\\
	&=
	\int_{X(0)=x_0}^{X(T)=x_T} DX(t) \, e^{- \frac{S[X(t)]}{\hbar}}  \, \hbar \, \delta^i_j \, \nonumber\\
	&=
	\langle x_T, T \,| \, \hbar \, \delta^i_j \, |\, x_0, 0 \rangle \, .
\end{align}
Therefore, we obtain the Euclidean commutator
\begin{equation}
	[\hat{x}^i,\hat{p}_j] = \hbar \, \delta^i_j \,,
\end{equation}
which, once Wick rotated back, reproduces the standard commutator $[\hat{x}^i,\hat{p}_j] = \ri \, \hbar \, \delta^i_j$.

\subsection{Spacetime non-commutativity due to discontinuous paths}\label{sec:NonContPath}
In the first subsection \ref{sec:Ordering}, we recalled that commutation relations manifest themselves in a path integral formulation as a difference in time-ordering. Then, in the previous subsection \ref{sec:PhaseSpaceNC}, we used this fact to present a derivation of the canonical commutation relations in a path integral formulation. In particular, we saw that the non-commutativity of position and momentum operators is reflected in the fact that the path integral integrates over paths that are nowhere differentiable.

Let us now consider a trajectory $X$ that may no longer be continuous. We assume, however, that both the left and right limits of $X$ exist for every $\tau\in\mathcal{T}$, but might not be equal. Thus, we write
\begin{align}\label{xpxm}
	X_+(t) &= \lim_{\epsilon\rightarrow 0^+} X(t+\epsilon) \, ,\nonumber\\
	X_-(t) &= \lim_{\epsilon\rightarrow 0^+} X(t-\epsilon) \, 
\end{align}
and define
\begin{align}
	\Delta X(t) &= X_+(t) - X_-(t) \,\\
	X_b(t) &= b\, X_+(t) + (1-b) \, X_-(t) \label{eq:FamXb}
\end{align}
with $b\in[0,1]$. 
\par 

In analogy with eq.~\eqref{eq:OrderingCorrespondence}, we can consider a spacetime commutator and identify it with a time ordering difference of the paths, such that
\begin{align}\label{eq:TimeOrderDiffX}
	\langle x_T, T \, | \, [\hat{x}^i,\hat{x}^j](t) \, | \, x_0,0\rangle
	&=
	\lim_{\epsilon\rightarrow0^+} \int_{X(0)=x_0}^{X(T)=x_T} DX(t) \, e^{-\frac{S[X(t)]}{\hbar}} 
	\nonumber\\
	&\qquad 
	\times
	\left[ X^i(t+\epsilon) X^j(\tau-\epsilon) - X^j(\tau+\epsilon) X^i(t-\epsilon) \right]
	\nonumber\\
	&=
	\int_{X(0)=x_0}^{X(T)=x_T} DX(t) e^{-\frac{S[X(t)]}{\hbar}} \left[ X_+^i(t) \, X_-^j(t) - X_+^j(t) \, X_-^i(t)  \right]
	\nonumber\\
	&= 
	\int_{X(0)=x_0}^{X(T)=x_T} DX(t) \, e^{-\frac{S[X(t)]}{\hbar}} \left[ X_b^j(t) \, \Delta X^i(t) -  X_b^i(t) \, \Delta X^j(t) \right] .
\end{align}
We immediately see that the commutator vanishes if $X$ is continuous, while it may be non-vanishing, if $X$ is not continuous. We conclude that the (non-)commutativity of spacetime can be related to a (dis)continuity of the paths appearing in the path integral formulation.
\par 

We will now illustrate this with an explicit example. We set $b=0$, and consider a massive relativistic particle with a trajectory $X^\mu(\tau)$ in the Minkowski space $\R^{n,1}$ parameterized by the proper time $\tau$. After a Wick rotation, the transition amplitude \eqref{eq:TimeOrderDiffX} for this theory can be written as\footnote{We refer to Refs.~\cite{Reisenberger:2001pk,Giovannetti:2022pab,Kuipers:2023pzm,Kuipers:2023ibv} for a discussion of covariant formulations of relativistic quantum mechanics.}
\begin{equation}\label{eq:CommutatorSpacetime}
	\langle x_T \, | \, [\hat{x}^\mu,\hat{x}^\nu](\tau) \, | \, x_0\rangle
	=
	\int_{X(0)=x_0}^{X(T)=x_T} DX(\tau) \, e^{-\frac{S[X(t)]}{\hbar}} \left[ X_-^\nu(\tau) \, \Delta X^\mu(\tau) -  X_-^\mu(\tau) \, \Delta X^\nu(\tau) \right] ,
\end{equation}
where $x_0,\,x_T\in\R^{n,1}$ are events such that $x_T$ lies in the future lightcone of $x_0$. Next, we assume that the spatial trajectories are continuous as a function of the proper time $\tau$. Hence, 
\begin{equation}\label{eq:Example}
	X_+^i = X_-^i \qquad {\rm and} \qquad \Delta X^i=0 \, ,
\end{equation}
which, by eq.~\eqref{eq:CommutatorSpacetime}, implies the commutation relations
\begin{equation}
	[\hat{x}^0,\hat{x}^0] = 0 \qquad {\rm and} \qquad [\hat{x}^i,\hat{x}^j] = 0 \, .
\end{equation}
Furthermore, we assume the existence of a minimal time scale $\Delta X^0 = \frac{1}{\kappa}$, such that any observer measures time with respect to a discrete clock. The time trajectory is then given by
\begin{align}
	X_-^0(\tau) &= \frac{1}{\kappa} \, \sum_{\tau_k<\tau} 1 \, ,\nonumber\\
	X_+^0(\tau) &= \frac{1}{\kappa} \, \sum_{\tau_k\leq\tau} 1 \, ,
\end{align}
where $\tau_k\in\{\tau_1,\tau_2,...\}\subset\mathcal{T}$ labels a finite set of `jump times' at which the clock ticks. We assume that these ticks arrive according to a Poisson distribution with rate $\nu(\tau)$. The number of ticks in an interval $[\tau,\tau+d\tau)$ is then given by
\begin{equation}
	X(\tau+d\tau) - X(\tau) 
	= \int_\tau^{\tau+d\tau} \rd X^0(s) \, 
\end{equation}
and the expected number of ticks in this interval is determined by the rate, such that
\begin{equation}\label{eq:dX0dtau}
	\langle X(\tau+d\tau)-X(\tau)\rangle
	= \int_{\tau}^{\tau+d\tau} \frac{\nu(s)}{\kappa} \, ds \, .
\end{equation}
%
%
%
\par

Let us now consider the commutator $[x^0,x^i]$. By eq.~\eqref{eq:CommutatorSpacetime}, we obtain
\begin{align}
	\langle x_T \, | \, [\hat{x}^0,\hat{x}^i](\tau) \, | \, x_0\rangle
	&=
	\int_{X(0)=x_0}^{X(T)=x_T} DX(\tau) \, e^{-\frac{S[X(t)]}{\hbar}} \left[ X_-^i(\tau) \, \Delta X^0(\tau) -  X_-^0(\tau) \, \Delta X^i(\tau) \right] \nonumber\\
	&= \int_{X(0)=x_0}^{X(T)=x_T} DX(\tau) \, e^{-\frac{S[X(t)]}{\hbar}} X_-^i(\tau) \, \Delta X^0(\tau) \, .
\end{align}
As in section \ref{sec:PIvsWP}, we will now integrate this expression along some arbitrary proper time interval $I\subseteq\mathcal{T}$, which yields
\begin{align}\label{eq:Der1}
	\left\langle x_T \left| \int_I [\hat{x}^0,\hat{x}^i](\tau) \, d\tau \right| x_0 \right\rangle
	&=
	\int_{X(0)=x_0}^{X(T)=x_T} DX(\tau) \, e^{-\frac{S[X(t)]}{\hbar}}
	\int_{I} X_-^i(\tau) \, \Delta X^0(\tau) \,  d\tau 
	\nonumber\\
	&=
	\int_{X(0)=x_0}^{X(T)=x_T} DX(\tau) \, e^{-\frac{S[X(t)]}{\hbar}}
	\kappa \int_{I} \frac{ X_-^i(\tau) \, \Delta X^0(\tau)}{\nu(\tau)} \,  \rd X^0(\tau)
	\nonumber\\
	&=
	\int_{X(0)=x_0}^{X(T)=x_T} DX(\tau) \, e^{-\frac{S[X(t)]}{\hbar}}
	\kappa \lim_{N\rightarrow\infty} \sum_{k=0}^{N-1} \frac{X_-^i(\tau_k)}{\nu(\tau_k)} \, [ X_-^0(\tau_{k+1}) - X_-^0(\tau_{k}) ]^2
\end{align}
where we used eq.~\eqref{eq:dX0dtau} in the second line and the definition of the It\^o integral \eqref{eq:ItoFDef} in the third line. Next, using that $X^0$ is a pure jump process with jump size $\kappa^{-1}$, one finds
\begin{equation}\label{eq:Jumphigherpow}
	[ X_-^0(\tau_{k+1}) - X_-^0(\tau_{k}) ]^2 = \frac{X_-^0(\tau_{k+1}) - X_-^0(\tau_{k})}{\kappa}
	= \begin{cases}
		\kappa^{-2} \quad & \textrm{if a jump occurs}\, ,\\
		0 \quad & \textrm{if no jump occurs}\, .
	\end{cases}
\end{equation}
Hence,
\begin{align}\label{eq:Der2}
	\left\langle x_T \left| \int_I [\hat{x}^0,\hat{x}^i](\tau) \, d\tau \right| x_0 \right\rangle
	&=
	\int_{X(0)=x_0}^{X(T)=x_T} DX(\tau) \, e^{-\frac{S[X(t)]}{\hbar}}
	\nonumber\\
	&\qquad\times
	\lim_{N\rightarrow\infty} \sum_{k=0}^{N-1} \frac{X_-^i(\tau_k)}{\nu(\tau_k)} \, [ X_-^0(\tau_{k+1}) - X_-^0(\tau_{k}) ] 
	\nonumber\\
	&= 
	\int_{X(0)=x_0}^{X(T)=x_T} DX(\tau) \, e^{-\frac{S[X(t)]}{\hbar}}
	\int_I \frac{X_-^i(\tau)}{\nu(\tau) } \, \rd X^0(\tau) 
	\nonumber\\
	&=
	\int_{X(0)=x_0}^{X(T)=x_T} DX(\tau) \, e^{-\frac{S[X(t)]}{\hbar}}
	\int_I \frac{ X_-^i(\tau) }{\kappa} \, d\tau \, .
\end{align}
Removing the integral on both sides yields 
\begin{equation}
	\left\langle x_T \left| [\hat{x}^0,\hat{x}^i](\tau) \right| x_0 \right\rangle
	=
	\int_{X(0)=x_0}^{X(T)=x_T} DX(\tau) \, e^{-\frac{S[X(t)]}{\hbar}} \frac{ X_-^i(\tau) }{\kappa} \, ,
\end{equation}
which implies the Euclidean commutation relation
\begin{equation}
	[\hat{x}^0,\hat{x}^i] = \frac{\hat{x}^i}{\kappa} \, .
\end{equation}
Finally, after performing a Wick rotation, one obtains
\begin{equation}\label{eq:kMinComm}
	[\hat{x}^0,\hat{x}^i] = \ri \, \frac{\hat{x}^i}{\kappa} \, .
\end{equation}
This commutation relation reproduces a non-commutative space-time of Lie algebra type known in the literature as $\kappa$-Minkowski spacetime \cite{Majid:1994cy,Arzano:2021scz}. In the path integral approach we adopted above we showed that this type of non-commutativity is associated with a fundamental discreteness in the temporal evolution, whereas space remains continuous.
\par

The presence of a minimal scale in time strongly suggests a breaking of the Lorentz symmetry in the path integral measure, which could be regarded as a problematic feature, if this model is to be related to reality. We emphasize, however, that, since we recovered the $\kappa$-Minkowski model, we do not expect the theory to be Lorentz invariant. Instead, one expects the theory to be invariant under a deformed Lorentz symmetry, as will be discussed in the next section. Nevertheless, this still raises the question, how such a deformed Lorentz symmetry manifests itself in the discrete time model described above. Although, we cannot yet provide a definitive answer to this question, we can make a few observations that could help in answering this question.
\par 

First, we note that the discreteness only relates to the evolution of time, but not to spacetime itself. Thus, an observer may measure any time $x^0=t\in\R$, but the difference between two time measurements performed by the same observer will always be a multiple of $\kappa^{-1}$. In other words, a time measurement yields $t=t_0 + n/\kappa$, were $n\in\mathbb{Z}$ but $t_0,t\in\R$. This means that the theory is still defined on a smooth manifold, but the generator of time translations has a discrete spectrum \cite{Lizzi:2021dud}. 
Secondly, the model is formulated in an arbitrary coordinate system, which makes the discreteness in time universal. Thus, any observer will measure time in units $\kappa^{-1}$. However, the model can still account for time dilation, since the rate $\nu(\tau)$ at which a clock ticks is not universal.
\par 

We conclude this section by noting that in our example we fixed $b=0$, such that the trajectory is projected onto its left limit, i.e. $X=X_-$. A similar analysis\footnote{In this case, one must use the definition of the backward It\^o integral \eqref{eq:ItoBDef} in eqs. \eqref{eq:Der1} and \eqref{eq:Der2}, in order to ensure convergence of the It\^o sums that define the integral, as the integrand needs to be predictable with respect to the integrator, cf. footnote \ref{footnote}. We refer to e.g. Refs.~\cite{RogersWilliams,KaratzasSchreve} for more detail.} can be performed for $b=1$, so that the trajectory is projected onto its right limit, i.e. $X=X_+$, which yields the same commutation relation \eqref{eq:kMinComm}. Furthermore, by linear interpolation, one obtains the commutation relation of the type \eqref{eq:kMinComm} for any choice of $b\in[0,1]$.

\section{Discontinuous paths and deformed Leibniz rule}\label{sec:Leibniz}
In the previous section, we used the relation between operator ordering in canonical quantization and time ordering in the path integral formalism to show that canonical non-commutativity between the position and momentum operator can be seen as an effect of the non-differentiability of the paths appearing in the path integral. We then made the observation that space-time non-commutativity can be realized for position operators by introducing discontinuity in the paths considered in the path integral. In this section, we show how another peculiar feature of models of space-time non-commutativity of the type \eqref{eq:kMinComm} can be realized introducing discontinuity in the paths appearing in the path integral: the non-trivial differential calculus and its relation to a deformation of the ordinary Leibniz rule \cite{Agostini:2006nc,Arzano:2021scz}. In particular we show that discontinous stochastic processes lead to differential operators which can reproduce the deformed Leibniz rule found in models of space-time non-commutativity in which four-momentum space is described by a non-abelian Lie group.

\subsection{Differentiable Trajectories}
We start by considering a classical trajectory $X:\mathcal{T}\rightarrow\R^n$ parameterized by the time variable $t\in\mathcal{T}=[0,T]\subset\R$, and smooth functions $f,g:\R^{n}\rightarrow\mathbb\R$. Along this trajectory, one can define differentials
\begin{align}\label{eq:DifferentialCont}
	\rd_+ X(t) &= X(t + dt) - X(t) \, ,\nonumber\\
	\rd_- X(t) &= X(t) - X(t-dt) \, 
\end{align}
with $0<dt\ll1$. This measures the change of the position with respect to an infinitesimal future and past directed time evolution. 
More generally, one can introduce a family of differential operators labeled by $a\in[0,1]$ given by
\begin{equation}\label{eq:DiffOpFam}
	\rd_a X(t) = X( t + a\, dt) - X( t - (1-a) \, dt) \, 
\end{equation}
with special cases
\begin{equation}
	\rd_+ = \rd_1 \, , \qquad \rd_- = \rd_0 \, . \qquad 
\end{equation}
Let us also introduce the notation
\begin{align}\label{eq:dsaplit}
	\rd_{a}^+ X(t) &= X(t+ a\, dt) - X(t) \, ,\nonumber\\
	\rd_{a}^- X(t) &= X(t) - X(t- (1-a)\,dt) \, ,
\end{align}
such that
\begin{equation}
	\rd_a X(t) = \rd_{a}^+X(t) + \rd_{a}^-X(t)\,.
\end{equation}
The differentials along the trajectory induce a family of differential operators $\rd_a$, whose action on scalar functions is given by
\begin{align}
	\rd_a f[X(t)] &= f[X(t+a\,dt)] - f[X( t - (1-a) \, dt) ] \, .
\end{align}
For smooth functions, this expression can be evaluated using a Taylor expansion, which yields
\begin{align}\label{eq:DifferentialFunctionCont}
	\rd_a f[X(t)] 
	&= \sum_{k_1=0}^\infty ... \sum_{k_n=0}^\infty \frac{\p_{1}^{k_1} ... \p_{n}^{k_n} f[X(t)]}{k_1! \, ... \, k_n!}
	\left\{ 
	\prod_{i=1}^{n} [\rd_a^+ X^i(t)]^{k_i}
	- \prod_{i=1}^{n} [ - \rd_a^- X^i(t)]^{k_i}
	\right\}  .
\end{align}
This can be written in shorthand notation as
\begin{align}\label{dpmshorthand}
	\rd_a f(X) &= \left(e^{\rd_a^+ X^i \, \p_i} - e^{-\rd_a^- X^i \, \p_i }\right) f(X) \, .
\end{align}
As in eq.~\eqref{eq:dsaplit}, we will also introduce
\begin{align}
	\rd_a^+ f(X) &= \left(e^{\rd_a^+ X^i \, \p_i} - 1\right) f(X)  \, ,\nonumber\\
	\rd_a^- f(X) &= \left(1 - e^{-\rd_a^- X^i \, \p_i }\right) f(X)  \, ,
\end{align}
such that
\begin{equation}
	\rd_a f(X) = \rd_a^+ f(X) + \rd_a^- f(X) \, .
\end{equation}
\par 

We will now assume that the trajectory $X(t)$ is continuously differentiable. This implies that\footnote{Here $o(dt)$ denotes the little-o notation, as introduced by Bachmann and Landau.}
\begin{equation}\label{eq:DiffDiff}
	\rd_a X(t) = \dot{X}(t) \, dt + o(dt) 
\end{equation}
and
\begin{align}\label{eq:d+d-diff}
	\rd_a^+ X(t) &= a \, \dot{X}(t) \, dt + o(dt) \, , \nonumber\\
	\rd_a^- X(t) &= (1-a) \, \dot{X}(t) \, dt + o(dt) \, .
\end{align}
Thus, in this case, the family of differentiable operators $\rd_a$ reduces to a single differential operator $\rd$, that acts on functions as
\begin{equation}\label{eq:DiffOpDiff}
	\rd f(X) = \p_i f(X) \, \rd X^i + o(dt) \, .
\end{equation}
From this last expression, one immediately finds that the differential operator $\rd$ satisfies the Leibniz rule
\begin{equation}\label{eq:DiffOpLeibniz}
	\rd(f \, g) = f\, \rd g + g\, \rd f + o(dt) \, .
\end{equation}

\subsection{Continuous Trajectories}\label{sec:ContLeibniz}
As discussed in section \ref{sec:PIvsWP}, the canonical commutation relations of quantum mechanics can be reproduced in the path integral formulation by promoting smooth trajectories to stochastic processes $X:\mathcal{T}\times \Omega\rightarrow\R^{n}$. Then, the path integral sums over the sample paths $X(\cdot,\omega):\mathcal{T}\times \Omega\rightarrow\R^{n}$ labeled by $\omega\in\Omega$. As in the classical theory, these sample paths are continuous. However, in contrast to the classical theory, they are nowhere differentiable. Hence, eq.~\eqref{eq:DiffDiff} is no longer true and this will lead to a generalization of eq.~\eqref{eq:DiffOpDiff} as we show below. These ``quantum trajectories" are, technically speaking, $\alpha$-H\"older continuous for $\alpha\in(0,\frac{1}{2})$, but fail to be $\alpha$-H\"older continuous for $\alpha\geq\frac{1}{2}$ \cite{Morters:2010}. Due to the importance of this observation for the argument that follows, we recall the definition of H\"older continuity
\begin{mydef} {\rm (H\"older Continuity)}
	A trajectory $X:\mathcal{T}\rightarrow\R^n$ is $\alpha$-H\"older continuous if there exists $C>0$ such that $||X(t)-X(s)||\leq C \, |t-s|^\alpha$ for all $s,t\in\mathcal{T}$.
\end{mydef}
Thus, in terms of Landau Big-O notation, $\alpha$-H\"older continuity with $\alpha\in(0,\frac{1}{2})$ implies that the differentials satisfy
\begin{equation}\label{eq:HolderBigO}
	\prod_{j=1}^k \rd_a X^{i_j} \sim O\big(dt^{k/2}\big) \qquad \forall\, i_j\in\{1,...,n\}, \; \forall \, k\in \mathbb{N}, \; \forall\, a\in[0,1]\, .
\end{equation}
Hence, along these trajectories eq.~\eqref{eq:DiffOpDiff} is modified to
\begin{equation}\label{eq:DiffOpConta}
	\rd_a f = \p_i f \, \rd_a X^i + \frac{1}{2} \, \p_j \p _i f \left(  \rd_a^+ X^i \rd_a^+ X^j - \rd_a^- X^i \rd_a^- X^j \right)  + o(dt)\, .
\end{equation}
This leads to a product rule of the form
\begin{equation}\label{eq:DiffOpContprod}
	\rd_a (f\,g) = f \, \rd_a g + \rd_a f \, g + \rd_a^+ f \, \rd_a^+ g - \rd_a^- f \, \rd_a^- g + o(dt)\, .
\end{equation}
\par 

This expression can be simplified by noticing that eq.~\eqref{eq:HolderBigO} implies that product of two differentials is linear in $dt$.\footnote{For the Wiener process $c_+$ is the diffusion constant associated to the future directed evolution, whereas $c_-$ is the diffusion constant associated to the past-directed evolution. In sec. \ref{sec:PhaseSpaceNC}, we saw that the choice $c_\pm=\hbar/m$ is consistent with the Euclidean path integral, whereas $c_\pm=\ri \, \hbar/m$ corresponds to the Lorentzian path integral.} Thus,
\begin{align}
	\rd_\pm X^i \rd_\pm X^j = c_\pm \, dt + o(dt) \, .
\end{align}
Due to this linearity, one obtains\footnote{This can be regarded as a generalization of eq.~\eqref{eq:d+d-diff} from $X\in\mathcal{C}^1$ to $X\in\mathcal{C}^{1/2}$.}
\begin{align}\label{eq:dxdxlin}
	\rd_a^+ X^i \rd_a^+ X^j &= a \, \rd_+ X^i \rd_+ X^j + o(dt) \, , \nonumber\\
	\rd_a^- X^i \rd_a^- X^j &= (1-a) \, \rd_- X^i \rd_- X^j + o(dt) \, .
\end{align}
As a consequence eqs.~\eqref{eq:DiffOpConta} and \eqref{eq:DiffOpContprod} are simplified to
\begin{align}
	\rd_a f &= \p_i f \, \rd_a X^i + \frac{1}{2} \, \p_j \p _i f \left( a \, \rd_+ X^i \rd_+ X^j + (a-1) \, \rd_- X^i \rd_- X^j \right)  + o(dt)\, ,\label{eq:DiffOpConta2} \\
	\rd_a (f\,g) &= f \, \rd_a g + \rd_a f \, g + a \, \rd_+ f \, \rd_+ g + (a-1) \, \rd_- f \, \rd_- g + o(dt) \, .\label{eq:DiffOpContprod2}
\end{align}

\par
In addition, we will assume
\begin{align}\label{eq:TimeRevC12}
	\rd_- X^i \rd_- X^j = \rd_+ X^i \rd_+ X^j + o(dt) \, .
\end{align}
Eqs.~\eqref{eq:DiffOpConta2} and \eqref{eq:DiffOpContprod2} can then be further simplified to
\begin{align}
	\rd_a f &= \p_i f \, \rd_a X^i + \frac{2 \, a - 1}{2} \, \p_j \p _i f(X) \, \rd_+ X^i \, \rd_+ X^j + o(dt)\, ,\label{eq:ItoCont}\\
	\rd_a(f \, g) &= f \, \rd_a g + \rd_a f \, g + (2\, a - 1) \, \rd_+ f \, \rd_+ g + o(dt)\, . 
\end{align}
\par 
We see that, in general, the Lebiniz rule \eqref{eq:DiffOpLeibniz} is modified when considering non-differentiable paths. However, there exists a special choice of differential operator,\footnote{In the study of path integrals, various choices are studied. The choice $a=1$ is related to the prepoint discretization of the path integral, the choice $a=1/2$ is to the midpoint discretization and the case $a=0$ is to the postpoint discretization, whereas, in stochastic analysis, the choice $a=1$ is used in It\^o calculus, $a=1/2$ in Stratonovich calculus, and $a=0$ in It\^o calculus after time reversal, cf. e.g Ref.~\cite{Chaichian:2001cz} for more detail.} given by $a=\frac{1}{2}$. In this case $\rd_\circ=\rd_{1/2}$ satisfies the usual rules eqs.~\eqref{eq:DiffOpDiff} and \eqref{eq:DiffOpLeibniz}, i.e.
\begin{align}\label{eq:Leibnizdcirc}
	\rd_\circ f &= \p_i f \, \rd_\circ X^i \, ,\nonumber\\
	\rd_\circ (f\,g) &= f \, \rd_\circ g + \rd_\circ f \, g\, .
\end{align}
One can define a differential operator in terms of a momentum operator $\hat{p}$ and an infinitesimal spatial displacement $dx$ as
\begin{equation}\label{eq:DiffOpform}
	\rd f = \ri \, \hat{p}_i (f) \, dx^i \, .
\end{equation}
It is then easy to see that the two formulations coincide, when we identify
\begin{align}
	\rd &= \rd_\circ \, , \\
	dx^i &=  \rd_\circ X^i\, ,\\
	\hat{p}_i &= - \ri \, \p_i \, ,
\end{align}
where $\hat{p}$ obeys the ordinary Leibniz rule
\begin{equation}
	\hat{p}_i(f\,g) = \hat{p}_i(f) \, g + f \, \hat{p}_i (g) \, .
\end{equation}
\par

We conclude this subsection by pointing out that the discussion above suggests the existence of an entire family of momenta labeled by $a\in[0,1]$. However, only one choice, namely $a=1/2$, turned out to be compatible with the usual definition of the momentum operator in a quantum theory. This raises a question whether one can associate different momentum operators to the differential $\rd_a$ for $a\neq1/2$. In principle, one would expect this to be true, but such operators do not necessarily need to be self-adjoint. This leaves the question whether the different choices of $\rd_a$ lead to physically distinct notions of momentum. 
The question about the physicality of these different choices for the momentum is tied to the debate on the physical difference between It\^o and Stratonovich stochastic calculus, cf. e.g. \cite{Kampen:1981}. Our analysis does not provide any new insights in this debate.

\subsection{Discontinuous trajectories}
As in Section 2, we consider again a stochastic trajectory\footnote{Note that, while the trajectory is considered to be stochastic, the deterministic case can be easily obtained by considering a sample space $\Omega$ containing only one element.} $X:\mathcal{T}\times \Omega\rightarrow\R^{n}$, but we will no longer assume that the trajectory is continuous. More precisely, we will assume that the trajectory has {\it a countable number of isolated jump discontinuities}, such that the left and right limits, given by
\begin{equation}\label{xpm1}
	X_\pm(t) = \lim_{\epsilon\rightarrow 0} X(t\pm\epsilon) \, ,
\end{equation}
exist, but are not necessarily equal. Since the jumps are isolated, this process can be decomposed into a continuous part $\bar{X}$ and jumps $\Delta X$, such that
\begin{align}\label{eq:Decomp}
	X_+(t) &= \bar{X}(t) + \sum_{t_k\leq t} \Delta X(t_k) \, ,\nonumber\\
	X_-(t) &= \bar{X}(t) + \sum_{t_k<t} \Delta X(t_k)\, ,
\end{align}
where $t_k\in\mathcal{T}$ labels the jump times at which 
\begin{equation}
	\Delta X(t)= X_+(t) - X_-(t) \neq 0 \, .
\end{equation}
Furthermore, as in eq.~\eqref{eq:FamXb}, one can define a family of paths 
\begin{align}
	X_b(t)
	&= b\, X_+(t) + (1-b) \, X_-(t) \nonumber\\
	&= X_-(t) + b \, \Delta X(t)
\end{align}
with $b\in[0,1]$. We can also define differentials, as in eqs. \eqref{eq:DiffOpFam}, but those will now also contain the jumps, and are given by \cite{Kallenberg:2021,He:1992}
\begin{align}
	\rd_a X(t)
	&= X[ t + a\, dt] - X[ t - (1-a) \, dt]  \nonumber\\
	&= \bar{X}[ t + a\, dt] - \bar{X}[ t - (1-a) \, dt] + \Delta X(t) \nonumber\\
	&= \rd_a \bar{X}(t) + \Delta X(t) \, ,
\end{align}
where we used that the jumps are isolated in the second line.\footnote{The fact that the jumps are isolated ensures that for small enough $dt$ at most one jump is contained in the interval $(t-dt,t+dt)$. This does not guarantee that a jump occurs in this interval, but if no jump occurs one simply has $\Delta X(t)=0$.}
The action of differentials on smooth functions $f,g\in\mathcal{C}^\infty(\R^n)$ is given by eq.~\eqref{dpmshorthand}:
\begin{align}\label{eq:dadisc}
	\rd_a f(X) 
	&= \left( e^{ \rd_a^+ X^i  \p_i}  - e^{ -\rd_a^- X^i \, \p_i } \right) f(X) \\
	&= \left( e^{ [\rd_a^+ \bar{X}^i + (1-b) \, \Delta X^i] \p_i}  - e^{ - [ \rd_a^- \bar{X}^i + b \, \Delta X^i ] \p_i} \right) f(X)  \, . \nonumber
\end{align}
Let us first consider the case of a purely discrete process, i.e. $\bar{X}^i(t)=0$, such that $\rd_a^\pm\bar{X}^i=0$. In this case, one finds
\begin{align}
	\rd_a f(X) 
	= \Delta f(X)
	:= f(X_+) - f(X_-) \, .
\end{align}
This leads to a product rule given by
\begin{align}
	\rd_a (f\,g) 
	&=
	\Delta(f\,g) \nonumber\\
	&= f(X_\pm) \, \Delta g + \Delta f \, g(X_\mp)
	\nonumber\\
	&=
	\theta_\pm f \, \rd_a g 
	+ \rd_a f \, \theta_\mp g \, ,\label{eq:LeibnizDisc}
\end{align}
where we introduced the projection operators, defined by
\begin{equation}
	\theta_\pm f(X) := f(X_\pm) \, ,
\end{equation}
as discussed in Ref.~\cite{Biane:2010}. We thus observe that, as pointed out in the previous section, for continuous trajectories the differential operator $\rd_\circ$ presents a unique choice among all differential operators $\rd_a$ with $a\in[0,1]$ for which the Leibniz rule is satisfied. However, for discrete trajectories, the presence of the projection operators $\theta_\pm$ spoils the Leibniz rule for the differential operator $\rd_\circ$ as well.  This violation of the Leibniz rule can be interpreted as 
a modification of the action of momentum operators. Indeed, introducing these operators as in eq.~\eqref{eq:DiffOpform}, one finds that their action on products of functions is given by
\begin{align}
	\hat{p}_i (f\,g)
	&=
	\theta_\pm (f) \, \hat{p}_i(g) + \hat{p}_i(f) \, \theta_\mp(g) \, \nonumber\\
	&= 
	\begin{cases}
		e^{(1-b)\Delta X^j\p_j} (f) \, \hat{p}_i(g) + \hat{p}_i(f) \, e^{-b\Delta X^j\p_j}(g) \, ,\\
		e^{-b\Delta X^j\p_j} (f) \, \hat{p}_i(g) + \hat{p}_i(f) \, e^{(1-b)\Delta X^j\p_j}(g) \, ,
	\end{cases}
\end{align}
where the choice for the upper or lower solution can be regarded as a time ordering prescription. As discussed in more detail below, departures from the Leibniz rule of this type for momentum operators are the hallmark of deformations of spacetime symmetries based on the theory of Hopf algebras and quantum groups \cite{Arzano:2021scz,Arzano:2021hpg} which are related to the type of space-time non-commutativity discussed in Section 2, i.e. the so-called $\kappa$-Minkowski space-time \cite{Majid:1994cy}.

In the remainder, we try to establish a direct connection with the type of modified Leibniz action related to $\kappa$-space-time non-commutativity. To this end, we consider a massive particle in Minkowski space-time\footnote{As in section \ref{sec:NonContPath}, we will Wick rotate, such that the analysis can be performed in a Euclidean setting.} $\R^{n,1}$, whose trajectory $X^{\mu}(\tau)$ is parameterized by the proper time $\tau$. 
Moreover, we will assume that $X^0(\tau)$ is a purely discrete trajectory with constant jump size, such that 
\begin{align}\label{eq:X0def}
	\bar{X}^0(\tau) &=0 \qquad \forall \tau\in\mathcal{T} \, ,\nonumber\\
	\Delta{X}^0(\tau) &= 
	\begin{cases} 
		\kappa^{-1} \quad &\textrm{if a jump occurs at time } \tau \,, \\
		0 \quad &\textrm{if no jump occurs at time } \tau \, ,
	\end{cases}
\end{align}
whereas $X^i$ is assumed to be $\alpha$-H\"older continuous for $\alpha\in(0,\frac{1}{2})$, such that
\begin{alignat}{2}
	\Delta{X}^i(\tau) &= 0 \qquad &&\forall \tau\in\mathcal{T}\, ,\nonumber\\
	\bar{X}^i(\tau) &= X^i(\tau) \qquad &&\forall \tau\in\mathcal{T} \, ,
\end{alignat}
and we assume that eqs. \eqref{eq:dxdxlin} and \eqref{eq:TimeRevC12} hold for $\bar{X}$.
Let us, furthermore, define the momentum operator in the temporal direction as usual:
\begin{equation}\label{eq:p0def}
	\hat{p}_0 = - \alpha \, \frac{\p}{\p x^0} \, ,
\end{equation}
where $\alpha=\ri$ in the Lorentzian theory, $\alpha=1$ in the Euclidean case and $\p/\p x^0$ is a derivative with respect to the time coordinate\footnote{Note that $\p/\p x^0$ acts on functions $f,g:\R^{n,1}\rightarrow\mathbb{C}$. These functions may be evaluated along the process $X:\mathcal{T}\rightarrow \R^{n,1}$, but derivatives are defined with respect to the coordinates on $\R^{n,1}$.}. The action of this operator on a product of functions is given by
\begin{equation}
	\hat{p}_0(f\,g) = \hat{p}_0(f) \, g + f \, \hat{p}_0(g) \, .
\end{equation}
However, as indicated by eq.~\eqref{eq:LeibnizDisc}, the spatial momentum operators no longer satisfy the usual coproduct. If we apply eq.~\eqref{eq:LeibnizDisc} directly to this example, we obtain
\begin{align}
	\hat{p}_i (f\,g)
	&=
	\theta_\pm (f) \, \hat{p}_i(g) + \hat{p}_i(f) \, \theta_\mp(g) \, \nonumber\\
	&= 
	\begin{cases}
		\exp\left[  \frac{(b-1) \, \Delta X^0}{\alpha} \, \hat{p}_0 \right] (f) \, \hat{p}_i(g) 
		+ \hat{p}_i(f) \, \exp\left[ \frac{b\,\Delta X^0}{\alpha}\, \hat{p}_0\right](g) \, ,\\
		\exp\left[\frac{b \,\Delta X^0}{\alpha} \, \hat{p}_0\right] (f) \, \hat{p}_i(g) 
		+ \hat{p}_i(f) \, \exp\left[\frac{(b-1) \, \Delta X^0}{\alpha} \hat{p}_0\right](g) \, .
	\end{cases}
\end{align}
Thus, at the jump times\footnote{In between jumps $X^0$ is constant, such that there is no dynamics as there is no change with respect to the time coordinate. Therefore, we are only interested in evaluating this expression when a jump occurs.}, one obtains\footnote{As in section \ref{sec:NonContPath}, we rotate $\kappa \rightarrow - \ri \, \kappa$ in the complex plane when transferring from the Euclidean to the Lorentzian theory.}
\begin{align}\label{eq:picoprod}
	\hat{p}_i (f\,g)
	&= 
	\begin{cases}
		\exp\left[  \frac{(b-1) \,}{\kappa} \, \hat{p}_0 \right] (f) \, \hat{p}_i(g) 
		+ \hat{p}_i(f) \, \exp\left[ \frac{b}{\kappa}\, \hat{p}_0\right](g) \, ,\\
		\exp\left[\frac{b}{\kappa} \, \hat{p}_0\right] (f) \, \hat{p}_i(g) 
		+ \hat{p}_i(f) \, \exp\left[\frac{(b-1) }{\kappa} \hat{p}_0\right](g) \, .
	\end{cases}
\end{align}
As discussed in detail for example in \cite{Arzano:2021scz}, the action on product of functions of the operators $\hat{p}_{\mu}$ is related to the description of tensor product of two representations of the Lie algebra of generators of translations in Minkowksi space-time and of the Poincar\'e algebra. Formally this is encoded in a mathematical object known as {\it co-product}, which, for ordinary quantum mechanical translation generators obeying the Leibniz rule when acting on products of functions, has the simple form
\begin{equation}
	\Delta \hat{p}_{\mu} = \hat{p}_{\mu} \otimes 1 + 1 \otimes \hat{p}_{\mu}\,.
\end{equation}
In our case, for $\hat{p}_{0}$ we still have an ordinary co-product
\begin{equation}\label{cpp0}
	\Delta \hat{p}_{0} = \hat{p}_{0} \otimes 1 + 1 \otimes \hat{p}_{0}\,,
\end{equation}
while for the generators of spatial translations we have from equation  \eqref{eq:picoprod},
\begin{align}\label{eq:picoprod2}
	\Delta \hat{p}_i 
	&= 
	\begin{cases}
		\hat{p}_i \otimes \exp\left[ \frac{b}{\kappa}\, \hat{p}_0\right] +  \exp\left[  \frac{(b-1) \,}{\kappa} \, \hat{p}_0 \right] \otimes \hat{p}_i  \, ,\\
		\hat{p}_i \otimes \exp\left[\frac{(b-1) }{\kappa} \hat{p}_0\right] \, + \exp\left[\frac{b}{\kappa} \, \hat{p}_0\right] \otimes \hat{p}_i\,.
	\end{cases}
\end{align}
Notice how the second line of the equation above can be obtained simply by swapping the factors in the tensor product, thus, in what follows, we will focus on the first line. In the case $b=0$ the first line of \eqref{eq:picoprod2} above becomes 
\begin{equation}
	\Delta \hat{p}_i  = \hat{p}_i \otimes 1 +  \exp\left[ - \frac{\hat{p}_0 \,}{\kappa} \,  \right] \otimes \hat{p}_i \,,
\end{equation}
which the readers familiar with the literature on $\kappa$-Minkowski space-time will recognize as the co-product for generators of spatial translations of the $\kappa$-Poincar\'e algebra in the Majid-Ruegg or {\it bicrossproduct} basis \cite{Majid:1994cy}. In this basis the co-product for the generators of time translations is undeformed as is for our stochastic process in equation \eqref{cpp0}, and it can be shown that such co-products are associated to a choice of normal ordering needed to define the product of non-commutative plane waves on $\kappa$-Minkowski space-time where the time coordinate is always put to the right \cite{Agostini:2003vg}. For $b=\frac{1}{2}$ instead the first line of \eqref{eq:picoprod2} reads
\begin{equation}
	\Delta \hat{p}_i  = \hat{p}_i \otimes \exp\left[\frac{\hat{p}_0 \,}{2\kappa} \right] +  \exp\left[ - \frac{\hat{p}_0 \,}{2\kappa} \,  \right] \otimes \hat{p}_i \,,
\end{equation}
and, together with the undeformed co-product for $\hat{p}_0$ in \eqref{cpp0}, correspond to the so-called {\it standard basis} of the $\kappa$-Poincar\'e algebra \cite{Lukierski:1992dt} corresponding to a choice of ordering of non-commutative plane waves where the time coordinate is symmetrically placed both on the left and on the right of all the spatial coordinates \cite{Agostini:2003vg}.
\par

The fact that there exist various choices of $b\in[0,1]$ corresponding to different momentum operators $\hat{p}$ raises the question whether this choice leads to physically different models. This question is closely related to the physical intrepretation of the choice $a\in[0,1]$ for the differential operator $\rd_a$ that was addressed in previous subsection. In the literature on Hopf algebras, the common view is that different bases, thus different choices for $b$, of the kappa-Poincar\'e Hopf algebra correspond to different observables and thus different kinematical models.
\par 

The previous analysis yields the coproduct encountered in the $\kappa$-Minkowski model, making it consistent with results of section \ref{sec:NonContPath} for this specific example. However, this derivation relies on the direct application of eq.~\eqref{eq:LeibnizDisc}, which was derived for a pure jump process, whereas the spatial processes in our example are continuous. One may thus wonder whether eq.~\eqref{eq:LeibnizDisc} is still applicable in this case. In order to check this, one must re-evaluate eq.~\eqref{eq:dadisc} for our example. We will now show, for the illustrative case $b=0$, that this is indeed true. 
\par 

It is known in the literature, cf. e.g.~\cite{RogersWilliams,KaratzasSchreve,He:1992,Kallenberg:2021}, that for a smooth function $f$ the integral 
\begin{equation}\label{eq:StochInt}
	\int \rd_\circ f(X)
\end{equation}
is convergent\footnote{Convergence refers to the convergence of the It\^o sums \eqref{eq:ItoFDef} and \eqref{eq:ItoBDef} that define the integral.} if $X$ is a semi-martingale with respect to either its past (for a process evolving forward in time) or future (for a process evolving backward in time) states. If $b=0$, $X$ is a left-continuous Poisson process, which is a semi-martingale with respect to the future states. In this case, this integral is evaluated as
\begin{equation}\label{eq:StochInt2}
	\int \rd_\circ f(X) = \int \Delta f(X) + \int \p_i f(X_+) \, \rd_\circ \bar{X}^i \, ,
\end{equation}
where the integrand is evaluated on its right limit.\footnote{A convergence requirement is that the integrand is predictable with respect to $X$ When $X$ is evolved forward in time, predictability means that the value of $f[X(t)]$ can be predicted from the knowledge of all previous states $\{X(t):t\in[0,t)\}$. For continuous trajectories, this is always true. However, for discontinuous trajectories, it is only true for $f(X_-)$, as a jump may occur at time $t$, and no information about this jump is given by the previous states. On the other hand, when $X$ is evolved backward in time, as considered here, predictability means that the value of $f[X(t)]$ can be predicted from the knowledge of all future states $\{X(t):t\in(t,T)\}$. For discontinuous trajectories, it is only true for $f(X_+)$, as a jump may occur at time $t$, and no information about this jump is given by the future states. This enforces the projection on the right limit.\label{footnote}}. Hence, by imposing this convergence requirement, the differential constructed in eq.~\eqref{eq:dadisc} should be given by
\begin{equation}\label{eq:Ito1}
	\rd_\circ f(X) = \Delta f(X) + \p_i f(X_+) \, \rd_\circ \bar{X}^i \, ,
\end{equation}

%
%
%
%
We note that that in order to obtain this form for the differential, one must fix an ordering in prescription in eq.~\eqref{eq:dadisc} and allow for a non-trivial commutator of the form
\begin{equation}
	\left[e^{\rd_\circ^\pm \bar{X}^i\p_i}, e^{\Delta X^0\p_0} \right] \neq 0 \, .
\end{equation}
The non-triviality of this commutator is related to a violation of Fubini's theorem, reflecting non-commutativity of the integral operations:\footnote{This is again related to the fact that the integral only converges if the correct time ordering is applied: the integrand must be projected on the right limit, whereas the integral is taken along a left continuous process.}
\begin{equation}
	\int\int f(X^0,\bar{X}^i) \, \Delta X^0 \, \rd_\circ \bar{X}^i \neq \int\int f(X^0,\bar{X}^i) \, \rd_\circ \bar{X}^i \, \Delta X^0 
	\, .
\end{equation}
\par 

One may now generalize eq.~\eqref{eq:Ito1} to a product of functions by imposing the same time-ordering as in eq.~\eqref{eq:Ito1}. This yields
\begin{align}\label{eq:ItoProd1}
	\rd_\circ ( f\,g) 
	&= f(X_+) \, \Delta g + \Delta f \, g(X)  + \p_i f(X_+) \,\rd_\circ \bar{X}^i \, g(X) + f(X_+) \p_i g(X_+) \rd_\circ \bar{X}^i   \, ,\nonumber\\
	&= f(X_+) \, \rd_\circ g + \rd_\circ f \, g(X) \, ,
\end{align}
%
%
%
which coincides with the one obtained for purely discrete processes in eq.~\eqref{eq:LeibnizDisc}. Hence, the given argument suggests that, for $b=0$, eq.~\eqref{eq:picoprod} remains valid in the given example, such that 
\begin{align}
	\hat{p}_i (f\,g)
	&= 
	e^{-\hat{p}_0/\kappa} (f) \, \hat{p}_i(g) 
	+ \hat{p}_i(f) \, g \, .
\end{align}

\section{Conclusions}\label{Conclusion}

We have argued that spacetime non-commutativity is inherently linked to a fundamental discreteness in the evolution of physical systems on spacetime. We started, in section \ref{sec:PathIntegral}, by reviewing Feynman's argument \cite{Feynman:1948ur} that the non-commutativity of quantum mechanics manifests in the path integral formulation through the non-differentiability of the paths that the path integral sums over. We then repeated this argument for non-continuous paths, and showed in eq.~\eqref{eq:CommutatorSpacetime} that this discreteness can be associated to a non-commutativity of spacetime coordinates. 
\par 

In section \ref{sec:Leibniz}, we provided an additional argument in support of the relation between discreteness and spacetime non-commutativity by showing that the study of discrete trajectories induces a modification of the Leibniz rule as in eq.~\eqref{eq:LeibnizDisc}. This modification can be related to the deformed Leibniz rule obtained in models of spacetime non-commutativity in which the four-momentum is described by a non-Abelian Lie group.
\par

In both cases, we illustrated our results by focusing on a model with a discrete time coordinate and continuous space coordinates. Here, we assumed that the dynamics in the time coordinate was governed by a Poisson process with jump size $\kappa$. In this example, we were able to mimic the properties of the $\kappa$-Minkowski non-commutative space-time. In section \ref{sec:PathIntegral}, we saw that the non-commutativity relation of this model could be obtained in this specific example. Then, in section \ref{sec:Leibniz} we saw that various deformations of the Leibniz rule could be obtained depending on whether the trajectory was assumed to be left- or right-continuous. We then pointed out that this choice for the left or right limit corresponds to a choice of basis for the $\kappa$-Poincar\'e algebra.
\par 

We note that, as shown by eqs.~\eqref{eq:CommutatorSpacetime} and \eqref{eq:LeibnizDisc}, the non-commutative features that were obtained in this work arise due to the discreteness of the model. In the examples, the discreteness was imposed in a stochastic way using a Poisson process, as this provides a natural way to implement the required discreteness. Moreover, from the stochastic analysis perspective, it has the advantage of being a semi-martingale, such that one can integrate along the process, with a quadratic variation given by the suggestive structure relation $\rd[x,x]=\kappa^{-1} \rd x$. It remains an open question, however, whether the implementation using a Poisson process is essential for recovering the models of spacetime non-commutativity studied in the literature. 
\par 

The implementation of the discreteness using a stochastic model has the advantage that the discreteness, and thus the non-commutativity, may appear for any value of the proper time $\tau$, whereas a deterministic implementation would predetermine the times $\tau_k$ at which the discreteness occurs. As the dynamics is continuous in between these points, the non-commutativity only occurs at these predetermined points. Moreover, the choice for a Poisson process, may not only be argued by the fact that it is one of the most elementary discrete-time stochastic processes, but also by the fact that the Poisson process allows for a Lorentz invariant embedding in the spacetime, as pointed out within the causal sets program \cite{Bombelli:1987aa,Surya:2019ndm}.
\par

In relation to the exploration of different discrete models, one may wonder whether other types of spacetime non-commutativity can be obtained using similar methods. We emphasize that the analysis in this work, in particular Eq.~\eqref{eq:CommutatorSpacetime}, implies a Lie algebra valued type of non-commutativity, where commutators are of the form $[x^i, x^j ] = C^{ij}_{k} x^k$. This leaves the question whether the argument can be generalized to obtain other types of spacetime non-commutativity, such as the Moyal type with commutator $[x^i, x^j ] = B^{ij}$. A starting point for such a study could be the observation that the Heisenberg commutation relations can be related to a discontinuity of the paths of a Wiener process when analyzed in phase space. It is important to keep in mind, however, that we arrived at the Lie algebra valued type of non-commutativity by implicitly assuming that the discreteness of the paths was such that it satisfies the H\"older condition, cf. section \ref{sec:ContLeibniz}, for all $\alpha\leq0$. This distinguishes our analysis from a phase space formulation of the paths of the Wiener process. In the latter case, the canonical type $[x^i, p_j ] = A^i_j$ is obtained, but the paths associated to the position have H\"older parameter $\alpha<1/2$, whereas the paths associated to momentum have holder parameter $\alpha<-1/2$, cf. e.g. \cite{Friz:2020}. This suggests that the type of non-commutativity is related to the degree of regularity of the paths that are being studied. Clearly, more work is needed to make this observation precise, which we will leave for future work.
\par

The observation that the regularity of the paths is key to understanding the divergent structure of quantum theories in the path integral formulation is not new. For example, in Ref.~\cite{Koch:2014cma}, it was suggested that the divergences in quantum theories could be tamed by regularizing the paths in the path integral. Interestingly, this also led to modified uncertainty principles, which have also been associated to spacetime non-commutativity, e.g. \cite{Battisti:2008xy}. In contrast, in this work, we considered paths that have less regularity than the ones in the standard quantum mechanical path integral. The lack of regularity was then associated to a non-commutativity of spacetime, where the cut-off scale $\kappa$ corresponds to the size of the jumps in the paths. As discussed in the introduction, this cut-off scale could also be used to regularize quantum theories. 
\par

Our study opens several avenues for further investigation. For instance, exploring additional models of Lie-algebra type space-time non-commutativity, like the $\mathfrak{sl}(2,\mathbb{R})$ non-commutativity describing quantum particles coupled to $2+1$-dimensional gravity \cite{Arzano:2014ppa} or the so-called $\rho$-Minkowski non-commutative spacetime \cite{Fabiano:2023uhg}. In that case it is known that time has a discrete spectrum \cite{Lizzi:2021dud}, but it would be interesting to check if they admit a similar stochastic interpretation as presented in this work. Another important task would be to attempt at establishing an explicit dictionary between stochastic and non-commutative differential calculus, both in the $\kappa$-Minkowski case we considered in the present work \cite{Juric:2015jxa,Rosati:2021sdf} and in other relevant non-commutative models. Finally, the potential relevance of non-trivial coproducts for the generators of translations in models of deformed kinematics associated to polymer quantization \cite{Amelino-Camelia:2017utp} suggests that this popular model, used to study effective models of quantum gravitational effect, might also admit a description in terms of fundamental stochastic structure. 
\par 

More broadly, the argument presented in this work suggests that stochastic structures could provide rigorous mathematical tools to model space-time discreteness and thus provide valuable insights for a variety of approaches to quantum gravity.

\section*{Acknowledgements}
We are grateful to Kilian Hersent and the anonymous referee for various insightful comments on the manuscript.
The work of FK is supported by a postdoctoral fellowship of the Alexander von Humboldt foundation and a fellowship supplement of the Carl Friedrich von Siemens foundation. FK also acknowledges the support of INFN Napoli where part of the work was carried out. MA acknowledges support from the INFN Iniziativa Specifica QUAGRAP. The research of MA was also carried out in part in the frame of Programme STAR Plus, financially supported by the University of Napoli Federico II and Compagnia di San Paolo.

\appendix

\section{The Wiener process}\label{ap:Wiener}
In this appendix, we recall some properties of the Wiener process. Here, we focus on the properties that are essential for our discussion on phase space non-commutativity. For a more detailed discussion of the Wiener process we refer to textbooks such as \cite{KaratzasSchreve,Morters:2010,LeGall,Chaichian:2001cz,SchreveII}
\par 

We start by introducing the Wiener process as the limit of a scaled random walk. Thus, we will consider a discrete time parameter $k\in\mathbb{N}_0$, and we define a coin toss experiment with sample space $\Omega=\{H,T\}$, sigma algebra $\Sigma(\Omega)=\{\emptyset, \Omega,\{H\},\{T\}\}$ and a probability measure $\mathbb{P}:\Sigma(\Omega)\rightarrow [0,1]$ generated by $\mathbb{P}(\{H\}) = \frac{1}{2}$.
\par 

We introduce such a coin toss experiment for every $l\in\mathbb{N}$ and $i\in\{1,...,n\}$, and we assume that all these experiments are mutually independent. We can now introduce the random variables $J^i_l:\Omega^i_l\rightarrow \{-1,+1\}$, such that
\begin{equation}
	J^i_l = 
	\begin{cases}
		+1 \qquad {\rm if} \quad \omega_l^i = H \, ,\\
		-1 \qquad {\rm if} \quad \omega_l^i = T \, .
	\end{cases}
\end{equation}
In addition, we define the sum
\begin{equation}
	W^i_k = W^i_0 + \sum_{l=1}^k J^i_l \, ,
\end{equation}
where $W_0\in \mathbb{Z}^n$ denotes the initial position. Then, the collection of random variables $W=\{W_k \, | \, k \in \mathbb{N}_0\}$ denotes an $n$-dimensional random walk $W:\mathbb{N}_0\times\Omega\rightarrow\mathbb{Z}^n$ with sample space $\Omega= \Pi_{i=1}^n \Pi_{l\in\mathbb{N}} \Omega^i_l$.
\par 

In order to construct a continuum limit of this process, we define a continuous version, $W:\R_{+}\times\Omega\rightarrow\R^n$, of the discrete process by linear interpolation. Thus, we define
\begin{equation}
	W^i(t) = W^i_{\lfloor t \rfloor } + J^i_{\lceil t \rceil} (t - \lfloor t \rfloor ) \, .
\end{equation}
This allows to introduce the scaled random walk: for any $N\in\mathbb{N}$, we define
\begin{equation}
	X_{(N)}(t) = \sqrt{\frac{\alpha}{N}} \, W(N t) \, ,
\end{equation}
where $\alpha\in[0,\infty)$.
The limiting process
\begin{equation}
	X(t) = \lim_{N\rightarrow \infty} X_{(N)}(t)
\end{equation}
is a Wiener process with diffusion constant $\alpha$ and initial position $X(0)$.
\par 

The Wiener process itself is defined within this continuum limit. By definition, a continuous $\R^n$-valued continuous time stochastic process is a function $X:\R_+\times(\Omega,\Sigma,\mathbb{P})\rightarrow(\R^n,\mathcal{B}(\R^n)$, where $\R_+$ represents the time interval and $\mathcal{B}(\R^n)$ is the Borel sigma algebra on $\R^n$. Moreover, $(\Omega,\Sigma,\mathbb{P})$ is a probability space, such that for every $\omega\in\Omega$ $X(\cdot,\omega):\R_+\rightarrow\R^n$ is a continuous trajectory, and the probability for selecting this path is given by the probability measure $\mathbb{P}:\Sigma\rightarrow[0,1]$. Using this terminology, we can now provide the definition of the Wiener process.
\begin{mydef}{\rm (Wiener process a.k.a. Brownian Motion)}\\
	A stochastic process $X:\R_+\times(\Omega,\Sigma,\mathbb{P})\rightarrow(\R^n,\mathcal{B}(\R^n))$ is a Wiener process with diffusion constant $\alpha\in(0,\infty)$, if it is almost surely continuous and has independent normally distributed increments, i.e.
	\begin{itemize}
		\item $\mathbb{P}\big(\lim_{s\rightarrow t} ||X_s - X_t|| = 0\big)=1\,;$
		\item $\forall \, t_1<t_2<t_3<t_4 \in \R_+, \; (X_{t_4} - X_{t_3}) \indep (X_{t_2} - X_{t_1})\, ;$
		\item $\forall \ t_1<t_2\in\R_+, \; (X_{t_2}-X_{t_1})\sim \mathcal{N}(0,\alpha \, \delta^{ij} \, (t_2-t_1))\,,$
	\end{itemize}
	where $\mathcal{N}(\mu,\sigma^2)$ denotes the normal distribution with mean $\mu\in\R^n$ and covariance matrix $\sigma^2\in\R^{n\times n}$.
\end{mydef}
%
%
%
\par 

The proof that the scaled random walk converges to this process is a result of the central limit theorem. As this proof can be found in many textbooks on Brownian motion, e.g. \cite{SchreveII}, we will not discuss it in detail here.
We point out, however, that the scaled random walk, as discussed above, is not the only process that converges to the Wiener process. One can construct other discrete time processes that converge to a Wiener process in their respective continuum limits.
\par 

\subsection{Velocity}\label{ap:WPV}
Having introduced the Wiener process, we discuss the notion of velocity for this process. In this subsection, we will provide a heuristic overview, which suffices to support the arguments in section \ref{sec:PIvsWP}. For a more rigorous construction of the velocity of a Wiener process, which exists as a distribution in the H\"older space $\mathcal{C}^{-1/2}$, we refer to e.g. \cite{Friz:2020}.
\par 

We start by considering the scaled random walk, $X_{(N)}:\R_+\times\Omega\rightarrow\R^n$, for which we can define two different velocities, which we call the \textit{forward and backward velocities}:
\begin{align}\label{eq:DefVelWienerDisc}
	V^i_{(N)+}(t) = \lim_{\epsilon\rightarrow 0^+} \frac{X_{(N)}^i(t+\epsilon) - X_{(N)}^i(t)}{\epsilon} \, ,\nonumber\\
	V^i_{(N)-}(t) = \lim_{\epsilon\rightarrow 0^+} \frac{X_{(N)}^i(t) - X_{(N)}^i(t-\epsilon)}{\epsilon} \, .
\end{align}
Evaluating these expressions yields
\begin{equation}\label{eq:VelWienerDisc+}
	V^i_{(N)+}(t) = 
	\begin{cases}
		\sqrt{\alpha \, N} \, J^i_{\lceil N t \rceil} \qquad &{\rm if} \quad Nt \in \R_+ \setminus \mathbb{N}_0 \, ,\\
		\sqrt{\alpha \, N} \, J^i_{Nt+1} \qquad &{\rm if} \quad Nt \in \mathbb{N}_0
	\end{cases}
\end{equation}
and
\begin{equation}\label{eq:VelWienerDisc-}
	V^i_{(N)-}(t) = 
	\begin{cases}
		\sqrt{\alpha \, N} \, J^i_{\lceil Nt \rceil} \qquad &{\rm if} \quad Nt \in \R_+ \setminus \mathbb{N}_0 \, ,\\
		\sqrt{\alpha \, N} \, J^i_{Nt} \qquad &{\rm if} \quad Nt \in \mathbb{N}_0 \, .
	\end{cases}
\end{equation}
Let us make two remarks:
\begin{itemize}
	\item $\forall t\in \R_+\quad V_{(N)\pm}\in\{+\sqrt{\alpha \, N}, - \sqrt{\alpha \, N}\}$ ;
	\item The definition of velocity is unambiguous, i.e. $V_{(N)+}=V_{(N)-}$, on the open intervals $t\in\left(\frac{k}{N},\frac{k+1}{N}\right)$ for every $k\in\mathbb{N}_0$. However, it may be ambiguous at the isolated points $t\in\{\frac{k}{N} \, | \, k\in \mathbb{N}_0\}$.
\end{itemize}
Furthermore, we note that any linear combination of the form
\begin{equation}
	V_{(N)a} = a \, V_{(N)+} + (1-a) \, V_{(N)-}
\end{equation}
with $a\in[0,1]$ provides a notion of velocity for the scaled random walk. This is particularly true for the case $a=\frac{1}{2}$ which can be written as
\begin{equation}\label{eq:VStratDisc}
	V_{(N)\circ}(t) = \lim_{\epsilon\rightarrow 0^+} \frac{X_{(N)}(t+\epsilon) - X_{(N)}(t-\epsilon)}{2\,\epsilon} \, .
\end{equation}
\par 

In the continuum limit, one can use eq.~\eqref{eq:DefVelWienerDisc} to define a velocity of the form
\begin{align}
	V^i_{+}(t) = \lim_{\epsilon\rightarrow 0^+} \lim_{N\rightarrow\infty} \frac{X_{(N)}^i(t+\epsilon) - X_{(N)}^i(t)}{\epsilon} \, ,\nonumber\\
	V^i_{-}(t) = \lim_{\epsilon\rightarrow 0^+} \lim_{N\rightarrow\infty} \frac{X_{(N)}^i(t) - X_{(N)}^i(t-\epsilon)}{\epsilon} \, .
\end{align}
However, eqs.~\eqref{eq:VelWienerDisc+} and \eqref{eq:VelWienerDisc-} cannot be used, as these expressions only hold for finite $N$. Nevertheless, if we naively swap the limit, we can apply these results and find
\begin{equation}\label{eq:LimAbsval}
	\lim_{N\rightarrow\infty}|V_{(N)\pm}^i(t)| = \infty \, .
\end{equation}
Moreover, the size of the intervals on which the velocity is unambiguously defined tends to $0$, which implies that the set of isolated points at which the velocity is ambiguous, becomes dense in $\R_+$. These two facts reflect the difficulty in defining a velocity for a Wiener process. Moreover, they are the source of the phase-space non-commutativity generated by the Wiener process, as discussed in section \eqref{sec:PhaseSpaceNC}.
\par 

As suggested by this naive analysis, it is indeed true that there is no well-defined notion of velocity for the Wiener process. More precisely, the limits 
\begin{align}\label{eq:VelWiener}
	\lim_{\epsilon\rightarrow 0^+} \frac{X^i(t+\epsilon)- X^i(t)}{\epsilon}
	\qquad 
	{\rm and}
	\qquad 
	\lim_{\epsilon\rightarrow 0^+} \frac{X^i(t)- X^i(t-\epsilon)}{\epsilon}
\end{align}
are ill-defined for every $t\in\R_+$ \cite{Morters:2010}. However, by averaging over all paths, one can construct well-defined forward and backward velocity fields \cite{Nelson:1985}. These are given by the conditional expectations
\begin{align}\label{eq:velfield}
	v_+(x,t) =  \lim_{\epsilon\rightarrow 0^+} \E\left[\frac{X(t+\epsilon)- X(t)}{\epsilon}\, \Big| \, X(t) = x\right]  ,\nonumber\\
	v_-(x,t) = \lim_{\epsilon\rightarrow 0^+} \E\left[\frac{X(t)- X(t-\epsilon)}{\epsilon}\, \Big| \, X(t) = x\right] .
\end{align}
These fields are not equal to each other, which reflects the fact that for every\footnote{Up to a set $A\subset\Omega$ with vanishing measure, i.e. $\mathbb{P}(A)=0$.} $\omega\in\Omega$ the sample path $X(\cdot,\omega):\R_+\rightarrow\R^n$ of the Wiener process $X:\R_+\times(\Omega,\Sigma,\mathbb{P})\rightarrow(\R^n,\mathcal{B}(\R^n))$ is continuous for every $t\in\R_+$, but nowhere differentiable.
\par 

More generally, any linear combination of the form
\begin{equation}
	v_a = a \, v_+ + (1-a) \, v_-
\end{equation}
with $a\in[0,1]$ defines a velocity of the process. As in eq.~\eqref{eq:VStratDisc}, we denote the special case $a=1/2$ by 
\begin{equation}
	v_\circ = \frac{v_+ + v_-}{2} \, .
\end{equation}
\par

The existence of the two linearly independent velocity fields allows to construct two velocity processes in terms of an integral expression. This requires an It\^o integral: given a differential form $\omega\in T^\ast\R^n$, and a Wiener process $X$, the \textit{It\^o integral} over $\omega$ along $X$ is given by
\begin{equation}\label{eq:ItoFDef}
	\int_{0}^T \omega_i(X(t),t) \, \rd_+X^i(t)
	=
	\lim_{N\rightarrow \infty} \sum_{k=0}^{N-1}
	\omega_i(X(t_k),t_k) 
	\big[X^i(t_{k+1}) - X^i(t_k) \big],
\end{equation}
where $\{0=t_0<t_1<...<t_N=T\}$ is an arbitrary partition of $[0,T]$.
Similarly, the backward It\^o integral is given by
\begin{equation}\label{eq:ItoBDef}
	\int_{0}^T \omega_i(X(t),t) \, \rd_-X^i(t)
	=
	\lim_{N\rightarrow \infty} \sum_{k=1}^{N}
	\omega_i(X(t_k),t_k) 
	\big[X^i(t_{k}) - X^i(t_{k-1}) \big].
\end{equation}
Using the existence of the It\^o integral, one can now define velocity processes $V_\pm$ as processes satisfying
\begin{equation}\label{eq:IntDefVel}
	\E\left[\int_{0}^T \omega_i(X(t),t) \, V^i_\pm(t) \, dt \right] 
	= 
	\E\left[\int_{0}^T \omega_i(X(t),t) \, \rd_\pm X^i(t)\right].
\end{equation}
We summarize by noting that the velocity $V_\pm$ of the Wiener process $X(t)$, which was naively defined in eq. \eqref{eq:VelWiener}, is ill-defined, as reflected by eq.~\eqref{eq:LimAbsval}. Nevertheless, we have found various properties of this process:
\begin{itemize}
	\item Its first moment is given by eq.~\eqref{eq:velfield}: $\E[V_\pm(t)|X(t)]=v_\pm(X(t),t)$;
	\item it has an action with respect to integration as given in eq.~\eqref{eq:IntDefVel}.
\end{itemize}

\subsection{Quadratic Variation}\label{sec:QVar}
An important concept in the study of stochastic processes is the notion of quadratic variation. For a discrete time process $W:\mathbb{N}_0\times\Omega\rightarrow \R^n$ it is defined by the square bracket
\begin{equation}
	[W^i,W^j]_k := \sum_{l=1}^k (W^i_l - W^i_{l-1}) \, (W^j_l - W^j_{l-1}) \, .
\end{equation}
For the random walk this can be evaluated, yielding
\begin{equation}
	[W^i,W^j]_k = 
	\begin{cases}
		k \, \qquad &{\rm if} \quad i=j \, ,\\
		k-2\,L \qquad &{\rm if} \quad i\neq j \, ,
	\end{cases}
\end{equation}
where $L\sim{\rm Bin}(k,\frac{1}{2})$ is binomially distributed.
\par 

The definition can easily be extended to the scaled random walk $X_{(N)}$. Indeed, for every $t_k=\frac{k}{N}\in\R_+$ with $k\in\mathbb{N}_0$, we can define the quadratic variation as
\begin{equation}
	[X^i_{(N)},X^j_{(N)}](t_k) = \sum_{l=1}^{k} [X_{(N)}^i(t_l) - X_{(N)}^i(t_{l-1})] \, [X_{(N)}^j(t_l) - X_{(N)}^j(t_{l-1})] \, .
\end{equation}
This expression can be evaluated, which yields
\begin{equation}\label{eq:QVarSRW}
	[X^i_{(N)},X^j_{(N)}](t_k) = 
	\begin{cases}
		\alpha \, t_k \, \qquad &{\rm if} \quad i=j \, ,\\
		\alpha \, (t_k - \frac{2 \, L}{N}) \qquad &{\rm if} \quad i\neq j \, ,
	\end{cases}
\end{equation}
where $L\sim{\rm Bin}(k,\frac{1}{2})$ is again binomially distributed.
\par 

More generally, for a continuous time stochastic process $X:\R_+\times\Omega\rightarrow\R^n$, one can define, for every bilinear form $h \in T^2 (T^\ast\R^n)$, the integral over $h$ along the quadratic variation of $X$ as
\begin{align}\label{eq:ItoQVDef}
	\int_{0}^T h_{ij}(X(t),t) \, \rd[X^i,X^j](t)
	&=
	\lim_{N\rightarrow \infty} \sum_{k=0}^{N-1}
	h_{ij}(X(t_k),t_k) \times
	\nonumber\\
	&\qquad
	\big[X^i(t_{k+1}) - X^i(t_k) \big] \big[X^j(t_{k+1}) - X^j(t_k) \big],
\end{align}
where $\{0=t_0<t_1<...<t_N=T\}$ is an arbitrary partition of the interval $[0,T]$. Thus, for a continuous time stochastic process the quadratic variation is given by
\begin{equation}
	[X^i,X^j](t) = \int_0^t \rd[X^i,X^j](s) \, .
\end{equation}
For the Wiener process, one can calculate this quadratic variation by taking the continuum limit of the scaled random walk. Using eq.~\eqref{eq:QVarSRW}, it is easy to see that the diagonal elements converge to $\alpha \, t$, while the off-diagonal elements converge to the delta distribution, as $\lim_{N\rightarrow\infty}\mathcal{N}(0,\frac{t}{N})=\delta(0)$, by the central limit theorem. Therefore,
\begin{equation}
	[X^i,X^j](t) = \alpha \, \delta^{ij} \, t \, ,
\end{equation}
such that
\begin{equation}\label{eq:QVarWiener}
	\rd[X^i,X^j](t) = \alpha \, \delta^{ij} \, dt \, .
\end{equation}
We note that the last expression allows to define a second order velocity given by
\begin{align}\label{eq:Vel2nd}
	V_2^{ij} 
	=\frac{ \rd [X^i,X^j] }{dt} 
	=\lim_{\epsilon \rightarrow 0} \frac{ [X^i(t+\epsilon) - X^i(t)] [X^j (t+\epsilon) - X^j (t)] }{\epsilon}
	\, .
\end{align}
Furthermore, we note that the integral over quadratic variation can be expressed in terms of the It\^o integrals \eqref{eq:ItoFDef} and \eqref{eq:ItoBDef} as
\begin{align}\label{eq:QvarvsIto}
	\int \rd[X^i,X^j](t)
	&=
	\lim_{N\rightarrow \infty} \sum_{k=0}^{N-1}
	\big[X^i(t_{k+1}) - X^i(t_k) \big] \big[X^j(t_{k+1}) - X^j(t_k) \big]
	\nonumber\\
	&=
	\lim_{N\rightarrow \infty} \left\{ \sum_{k=1}^{N}
	X^i(t_k) \big[X^j(t_{k}) - X^j(t_{k-1}) \big]
	- \sum_{k=0}^{N-1}
	X^i(t_k) \big[X^j(t_{k+1}) - X^j(t_k) \big] \right\}
	\nonumber\\
	&=
	\int X^i(t) \, \rd_-X^j(t) - \int X^i(t)\, \rd_+X^j(t) \, .
\end{align}

\end{document}